\documentclass[aps,prl,twocolumn,groupedaddress,floatfix]{revtex4}

\usepackage{graphicx}
\usepackage{dcolumn} 
\usepackage{bm}      

\begin{document}

\title{The potential energy of a $^{40}$K Fermi gas in the BCS-BEC crossover}

\author{J. T. Stewart}
\email[Electronic address: ]{J.Stewart@jila.colorado.edu}
\homepage[URL: ]{http://jilawww.colorado.edu/~jin/}
\author{J. P. Gaebler}
\author{C. A. Regal}
\author{D. S. Jin}

\affiliation{JILA, Quantum Physics Division, National Institute of
Standards and Technology and Department of Physics, University of
Colorado, Boulder, CO 80309-0440, USA}

\date{\today}
\begin{abstract}
We present a measurement of the potential energy of an ultracold
trapped gas of $^{40}$K atoms in the BCS-BEC crossover and
investigate the temperature dependence of this energy at a wide
Feshbach resonance, where the gas is in the unitarity limit. In
particular, we study the ratio of the potential energy in the region
of the unitarity limit to that of a non-interacting gas, and in the
$T=0$ limit we extract the universal many-body parameter $\beta$. We
find $\beta = -0.54^{+0.05}_{-0.12}$; this value is consistent with
previous measurements using $^{6}$Li atoms and also with recent
theory and Monte Carlo calculations. This result demonstrates the
universality of ultracold Fermi gases in the strongly interacting
regime.
\end{abstract}

 \pacs{??)}

\maketitle

With the emergence of novel Fermi gas systems, experimentalists can
now access the Bardeen Cooper Schrieffer (BCS)- Bose-Einstein
Condensate (BEC) crossover in ultracold gases of atoms. Using atomic
scattering resonances in gases of $^{40}$K and $^6$Li, it is
possible to widely tune the interatomic scattering length, $a$, and
move continuously between a gas of weakly interacting fermions and a
gas of condensed molecules. The BCS-BEC crossover occurs in the
strongly interacting regime where the scattering length is large
enough that $-1 \lesssim 1/k_F a \lesssim 1$, where $k_F$ is the
Fermi wave vector. Experimental studies of these strongly
interacting Fermi systems have revealed many interesting properties
including a phase transition involving condensates of atom pairs
\cite{Regal2004a, Zwierlein2005a}, a pairing gap \cite{Chin2004a},
and vortices in a rotating gas \cite{Zwierlein2005b}.

As a gas of fermions is cooled from the classical regime to quantum
degeneracy, the Pauli exclusion principle becomes manifest in the
properties of the ultracold gas \cite{DeMarco1999a, Truscott2001a}.
For example, a zero-temperature Fermi gas in a confining potential
has a finite energy and a finite size due to Fermi pressure, which
is responsible for the stability of white dwarf and neutron stars.
As the two-body scattering length is tuned to be arbitrarily
attractive, one would expect that the gas should be compressed due
to attractive interactions and pairing effects \cite{Stoof1996a,
Gehm2003a}.

This behavior has received theoretical consideration and should not
depend on the details of the interatomic potential for a wide
Feshbach resonance, where the Fermi energy is much smaller than the
energy equivalent width of the resonance \cite{Diener2004a,
Diener2004b, Simonucci2005, Szymanska2005a}. Furthermore, at
resonance, where the two-body scattering length $a$ diverges, the
energy of the gas is expected to be universal \cite{Heiselberg2001a,
Baker1999a, Ho2004a} in that it depends only on the Fermi energy
$E_F$ and the relative temperature $T/T_F$. The density profile of a
trapped $T=0$ unitarity limited gas is then expected to be simply a
rescaled version of the non-interacting density profile. This
results in a simple rescaling of the size and energy, which can be
parameterized by a universal many-body parameter $\beta$
\cite{O'Hara2002a,Menotti2002}.

Experimentally, $\beta$ has only been reported for $^6$Li where the
size and energy of a trapped gas has been examined
\cite{O'Hara2002a,Gehm2003a, Bourdel2003a, Bartenstein2004a,
Bourdel2004a, Partridge2005a, Kinast2005a}. The most precise
determination, $\beta = -0.54\pm0.05$, was reported recently in Ref.
\cite{Partridge2005a}. While this value is in good agreement with
the predicted value, a measurement using a different atomic species
is essential to demonstrate the universality of strongly interacting
Fermi gases. Here we report on a measurement of the potential energy
at resonance for an ultracold gas of $^{40}$K. We have also measured
the potential energy throughout the strongly interacting regime and
investigated the temperature dependence of the potential energy at
resonance.

For these experiments we cool a gas of fermionic $^{40}$K atoms to
ultracold temperatures using previously described methods
\cite{Regal2005c}. A nearly equal mixture of the two lowest energy
hyperfine spin states, $|f,m_f \rangle = |9/2,-9/2 \rangle$ and
$|9/2,-7/2 \rangle$, is confined in a crossed beam optical dipole
trap. The trap consists of a horizontal laser beam parallel to
$\hat{z}$ with a $\frac{1}{e^2}$ radius of 32 $\mu$m and a vertical
beam parallel to $\hat{y}$ with a $\frac{1}{e^2}$ radius of 200
$\mu$m. For the experiments reported here the harmonic trap
frequencies were typically $\omega_r/2\pi=184$ Hz and
$\omega_z/2\pi=18$ Hz. Approximately $10^5$ atoms per spin state are
cooled to a final temperature of $T\approx 0.08 \hspace{0.05cm}T_F$,
where $T_F = E_F/k_b$ is the Fermi temperature and $k_b$ is
Boltzmann's constant.
\begin{figure}
\includegraphics[width=\linewidth]{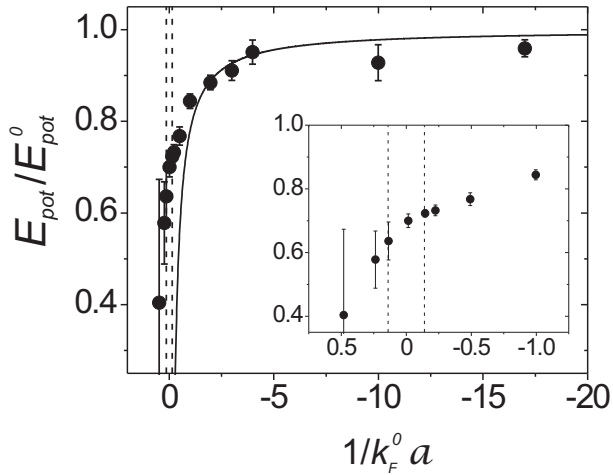}
\caption{\label{E_vs_kfa}Measured potential energy $E_{pot}$
normalized to the value measured in the non-interacting regime
$E_{pot}^0$ versus 1/$k_F^0 a$. Here ($^0$) denotes a quantity
measured in the non-interacting regime, i.e., at the zero crossing
of the s-wave scattering length. The resonance is located at
$202.10\pm0.07$ G \cite{Regal2004a}; the dashed lines show the
uncertainty in the resonance location. Data points toward the BCS
limit show good agreement with a zero temperature mean-field
calculation (solid line). The larger error bars on the BEC side of
the resonance reflect uncertainties due to heating of the gas due to
inelastic loss. In the strongly interacting region their exists
$\pm0.1$ uncertainty in $1/k_F^0 a$ due to uncertainty in the
resonance position. (inset) Subset of the data focusing on the
strongly interacting region near resonance.}
\end{figure}

The final evaporative cooling of the gas occurs on the BCS side of
the magnetic-field Feshbach resonance where $a\approx-1000
\hspace{0.07cm} a_0$. The optical trap is then ramped up to
approximately 1.5 times of the shallowest trap depth used for
evaporation. To vary $a$, the magnetic field is adiabatically ramped
to various final values. The optical trap is then suddenly switched
off and the gas is allowed to expand for 1.867 ms. During this short
expansion time there is significant expansion in the radial
direction but negligible expansion of the cloud in the axial
direction. We then use absorption imaging to probe the density
distribution of the atom cloud. The probe beam propagates along one
of the radial directions, $\hat{x}$, and is pulsed on for 40 $\mu$s.
For each absorption image we perform a 2D surface fit to a finite
temperature Fermi-Dirac function
\begin{equation}\label{eq:OD}
OD(y,z)=pk \hspace{0.07cm} g_2 \bigg(-\zeta  \hspace{0.05cm}
e^{-\frac{y^2}{2 \sigma_y^2}-\frac{z^2}{2 \sigma_z^2}}\bigg)/g_2
(-\zeta)
\end{equation}
where $\zeta$, $\sigma_{y}$, $\sigma_{z}$, and $pk$ are independent
fitting parameters and
$g_n(x)=\sum\limits_{k=1}^{\infty}\frac{x^k}{k^n}$
\cite{DeMarco2001c}. This is the expected optical depth (OD)
distribution for a non-interacting cloud both in trap and after
expansion. Empirically we find that this function also fits well in
the strongly interacting regime. The potential energy of the trapped
gas is obtained from the cloud profile in the axial direction.  The
potential energy per particle in the axial direction ($\hat{z}$) is
given by
\begin{equation}\label{eq:Energy}
{E_{pot}=\frac{1}{2} m  \hspace{0.05cm}\omega_z^2 \hspace{0.05cm}
\sigma_z^2 \hspace{0.05cm} \frac{g_4(-\zeta)}{g_3(-\zeta)}}
\end{equation}
where $m$ is the mass of $^{40}$K.

It is useful to normalize the measured potential energy of the
strongly interacting gas to that of an ideal (non-interacting) Fermi
gas. In previous experiments using $^6$Li atoms, the measured cloud
sizes and energies were normalized to a calculated value for the
non-interacting gas. This can introduce systematic errors because
the calculation relies on the atom number and trap frequencies,
which can have systematic errors. However, for the $^{40}$K Feshbach
resonance, we are able to reduce the systematic uncertainty by
measuring the potential energy of the non-interacting gas. This can
be accomplished by going to the zero crossing of the resonance where
$a=0$, which is only 10 G away. The measured potential energy ratio
for an ultracold $^{40}$K gas is displayed in Fig. \ref{E_vs_kfa} as
a function of 1/$k_F^0 a$. In this paper we use a superscript naught
$(^0)$ to denote measurements made in the non-interacting regime.
The inset to Fig. \ref{E_vs_kfa} focuses on the strongly interacting
region near the resonance. As expected, the data show that the
interactions cause a strong reduction in the potential energy due to
a compression of the trapped gas. One would expect that on the BEC
side of the resonance $E_{pot}$ will depend on condensate fraction.
For temperatures similar to these experiments we find a maximum
condensate fraction of approximately 15\% on resonance; this
fraction decreases as detuning from the resonance is increased
\cite{Regal2004a}.

\begin{figure}
\includegraphics[width=\linewidth]{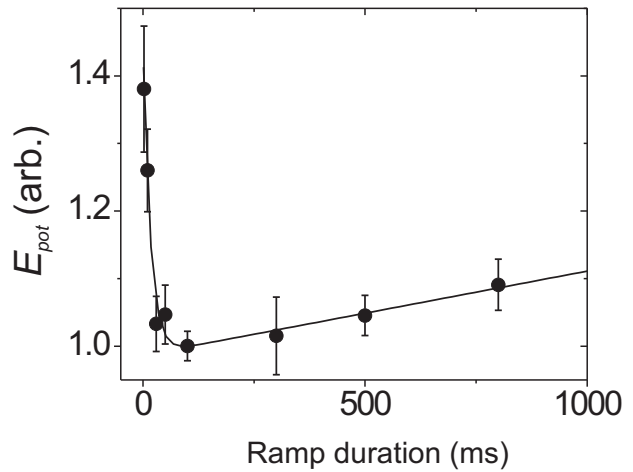}
\caption{\label{size_ramp_rate} Measured potential energy at the
Feshbach resonance versus magnetic-field ramp duration. For very
fast ramps, we measure a higher energy because of nonadiabaticity.
For very slow ramps, heating due to inelastic collisions increases
the measured energy.}
\end{figure}

The error bars in Fig. \ref{E_vs_kfa} include statistical
uncertainty in repeated measurements as well as an uncertainty due
to heating during the magnetic-field ramps. The magnetic-field ramps
must be sufficiently slow to be adiabatic; however, heating during
the ramp can be a problem for slower ramps. For the different final
magnetic fields we investigated the dependence of the measured
potential energy on the duration of the linear ramp. An example of
this is shown in Fig. \ref{size_ramp_rate} where the gas was ramped
from the magnetic field used for evaporation (203.39 $\pm$ 0.01 G)
to the resonance position.

To determine the optimum ramp rate, as well as the effect of heating
on the potential energy measurement, we fit data such as that shown
in Fig. \ref{size_ramp_rate} to an exponential decay plus linear
heating. From the fit we determine the final potential energy of the
cloud if heating were not present. This introduces a correction that
is applied to the data shown in Fig. \ref{E_vs_kfa}. Note that on
the BCS side of the resonance we see little or no heating due to
magnetic-field ramps, and the error bars are dominated by
shot-to-shot statistical uncertainty.

We can gain some theoretical insight into the effect of interactions
on the energy of our trapped gas by considering a simple mean-field
approach. While this approach neglects pairing and therefore is not
sufficient to fully understand the behavior of our gas, it provides
a flavor of how the potential energy is affected by interactions.
Following the argument outlined in Ref. \cite{Menotti2002}, the
equation of state for an ideal zero-temperature Fermi gas is
\begin{equation}\label{eq:EOS}
\mu =\epsilon _F (\mathbf x) +U_{MF}(\mathbf x) +U_{trap}(\mathbf
x),
\end{equation}
where $\mu$ is the chemical potential, $\epsilon _F (\mathbf x)$ is
the local Fermi energy,  $U_{MF}(\mathbf x)$ is the mean-field
contribution, and $U_{trap}(\mathbf x)$ is the trapping potential.
We can relate $\epsilon_F(\mathbf x) = \frac{\hbar^2}{2
m}k_F^2(\mathbf x)$ to the density $n(\mathbf x)=\frac{1}{6
\pi^2}k_F^3(\mathbf x)$ via $\epsilon_F(\mathbf x)=\frac{\hbar^2}{2
m}[6\pi^2n(\mathbf x)]^{2/3}$. The interactions appear in the
density dependent mean-field contribution, $U_{MF}(\mathbf
x)=\frac{4\pi\hbar^2 a}{m}n(\mathbf x)$.  This equation can be
solved self-consistently to determine the in-trap density profile of
the cloud, and thus the potential energy per atom.  In Fig.
\ref{E_vs_kfa} we compare the data to the mean-field calculation
(solid line) for the normal state on the BCS-side of the resonance
\cite{Regal2005c}.  Near the resonance, in the strongly interacting
regime, it is clear that this approximation breaks down and a more
sophisticated theory is required.

Very near the resonance the scattering length $a$ diverges and the
equation given above for $U_{MF}(\mathbf x)$ becomes unphysical.  At
resonance ($1/k_Fa=0$), the only energy scale is the Fermi energy;
this give us an approximate effective scattering length $a_{eff} =
-1/k_F$. This substitution shows that the local mean-field energy is
proportional to the Fermi energy, and one can define a constant of
proportionality $\beta$ given by $U_{MF}(\mathbf
x)=\beta\epsilon_F(\mathbf x)$ \cite{Gehm2003a,O'Hara2002a}. In this
simple mean-field estimate, $U_{MF}(\mathbf x)
=-\frac{4}{3\pi}\epsilon_F(\mathbf x)$, or $\beta_{MF}=-0.41$. The
negative sign for the scattering length is not obvious from this
approach, but a more sophisticated many-body approach shows the
mean-field interaction should be attractive \cite{Heiselberg2001a}.
Now we can write Eq. (\ref{eq:EOS}) as
\begin{equation}\label{eq:newEOS}
\mu =(1+\beta)\epsilon _F (\mathbf x) + U_{trap}(\mathbf x).
\end{equation}

Solving for the density profile for a harmonic trap and then
integrating to find the energy per particle, one finds that the
potential energy of a $T=0$ gas in the unitarity limit is simply
$E_{pot}=\frac{3}{8}\mu$, just as in the case of a non-interacting
Fermi gas.  To find the ratio of the chemical potential at resonance
$\mu$ to that for a non-interacting gas $\mu^0$, we hold the number
constant for each case, $N=\int n^0(x)d^3 x = \int n(x)d^3 x$, to
find
\begin{equation}\label{eq:murelation}
\frac{\mu}{\mu^0}=\sqrt{1 + \beta}.
\end{equation}
Thus, the universal parameter $\beta$ can be extracted by measuring
the ratio of the potential energy at resonance to the potential
energy of a non-interacting, trapped Fermi gas, in the $T=0$ limit.

\begin{figure}
\includegraphics[width=\linewidth]{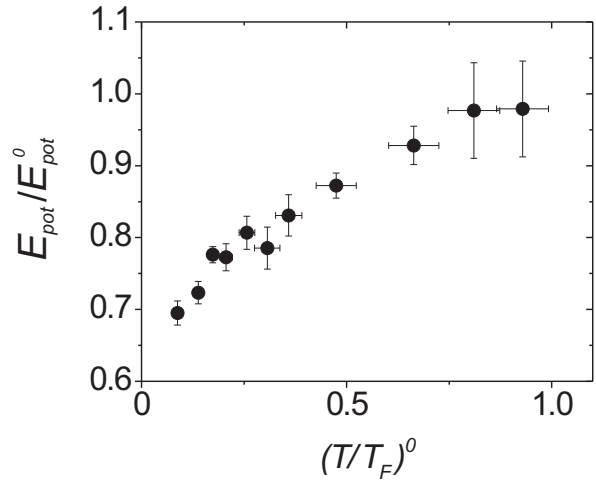}
\caption{\label{TTFparabola}Potential energy $E_{pot}$ normalized to
the measured energy in the non-interacting regime $E^0_{pot}$ vs the
non-interacting gas temperature $(T/T_F)^0$. The cloud is heated by
parametrically modulating the trapping potential. For these data the
trapping frequencies in the radial direction vary from $\sim$180 Hz
to 450 Hz and in the axial direction from $\sim$18 Hz to 21 Hz.}
\end{figure}

From the data in Fig. \ref{E_vs_kfa}, we find $E/E_0 = 0.70\pm0.02$
on resonance giving $\beta^*=-0.51\pm0.03$. Whereas $\beta$ is
normally only defined for $T=0$, we introduce ($^*$) to denote that
the system is at a finite temperature $(T/T_F)^0 = 0.08\pm0.01$.
This experiment was conducted at the Feshbach resonance and the
error bars include statistical error as well as the heating effects
mentioned above. Including uncertainty in the resonance position we
find $\beta^*=-0.51^{+0.04}_{-0.12}$.

We now consider experimentally the question of whether the gas is
sufficiently cold to be in the $T=0$ limit as required for an
accurate determination of $\beta$. In Fig. \ref{TTFparabola} we show
the measured potential energy ratio as a function of $(T/T_F)^0$. In
this experiment, we heat the gas by parametrically modulating the
optical trap strength. However, before heating the gas, we first
increase the optical trap depth to prevent number loss and ensure
harmonic confinement. Both the resulting temperature $(T/T_F)^0$ and
$k_F^0$ are determined using an ultracold cloud prepared as
described above and slowly ramping the magnetic field to the point
where $a=0$ just before the trap is turned off. The gas is allowed
to expand freely for 14 ms and fit according to Eq. (\ref{eq:OD}).
We extract $(T/T_F)^0$ from the fugacity $\zeta$ of the
non-interacting cloud using $g_2(-\zeta) = - (T/T_F)^{-3}/6$. It is
important to note that in this experiment we expect the
magnetic-field ramps to keep the entropy constant but not the
temperature.

The data in Fig. \ref{TTFparabola} clearly show that the universal
many-body parameter $\beta$ depends strongly on temperature.
Furthermore, it is not clear that a gas at $(T/T_F)^0 = 0.08$ is
sufficiently cold to determine the $T=0$ limit of $\beta$. For the
purpose of extrapolating to zero temperature, we fit a quadratic
function to the data points below $(T/T_F)^0$ = 0.25 for which we
find $\beta = -0.54^{+0.05}_{-0.12}$. The error bars reflect the
uncertainty in the extrapolation to $T=0$ and the uncertainty in the
resonance position. This value of the universal many-body parameter
$\beta$, as well as the value at $(T/T_F)^0$ = 0.08, is in good
agreement with Monte Carlo calculations \cite{Carlson2003a,
Carlson2005a, Astrakharchik2004a, Astrakharchik2005b, Burovski2006a}
and recent theoretical calculations \cite{Perali2004b, Hu2006a,
Hu2006b, Liu2005a}; in particular, we note $\beta=-0.545$ in Ref.
\cite{Perali2004b}. These values are also in good agreement with
multiple experimental reports in $^6$Li: $\beta =
-0.73^{+0.12}_{-0.09}$, $-0.61\pm0.15$, $-0.49$, and $-0.54\pm0.05$
in Refs. \cite{Bartenstein2004c}, \cite{Bourdel2004a},
\cite{Kinast2005a}, and \cite{Partridge2005a} respectively. We also
note that from the kinetic energy measurement in $^{40}$K reported
in Refs. \cite{Regal2005c, Chen2006a}, we can extract $\beta =
-0.62\pm0.07$ \cite{beta_kin}. This is in good agreement with the
value of $\beta$ found using the potential energy presented in this
paper.

In summary we have studied the potential energy of a strongly
interacting quantum degenerate gas of $^{40}$K Fermi atoms. At
resonance limit our results are consistent with current theory as
well as previous experiments in $^6$Li, thereby strengthening the
theory of universality of these Fermi gas systems. Universality
necessarily assumes that the s-wave Feshbach resonances in $^{40}$K
and $^6$Li are wide. The question of whether or not the $^{40}$K
Feshbach resonance is wide has been under debate
\cite{Szymanska2005a, Diener2004b, Mackie2005a, Javanainen2005a};
however, the good agreement of our results with $^6$Li provides
compelling evidence that $^{40}$K is also a wide resonance. We have
also measured the temperature dependence of a universal many-body
parameter that could be compared to recent Monte Carlo results for
the temperature dependent energy of a homogeneous Fermi gas
\cite{Burovski2006a, Bulgac2006a}.

\begin{acknowledgments}
We thank the JILA BEC group for stimulating discussions. This work
was supported by the NSF, NIST, and NASA; C.\,A.\,R. acknowledges
support from the Hertz Foundation.
\end{acknowledgments}



\begin{thebibliography}{10}

\bibitem{Regal2004a}
C.~A. Regal, M. Greiner, and D.~S. Jin, Phys. Rev. Lett. {\bf 92},
040403
  (2004).

\bibitem{Zwierlein2005a}
M.~W. Zwierlein {\it et~al.}, Phys. Rev. Lett. {\bf 92},  120403
(2004).

\bibitem{Chin2004a}
C. Chin {\it et~al.}, Science {\bf 305},  1128  (2004).

\bibitem{Zwierlein2005b}
M.~W. Zwierlein, J.~R. Abo-Shaeer, A. Schirotzek, C.~H. Schunck, and
W.
  Ketterle, Nature {\bf 435},  1047  (2005).

\bibitem{DeMarco1999a}
B. DeMarco and D.~S. Jin, Science {\bf 285},  1703  (1999).

\bibitem{Truscott2001a}
A.~G. Truscott, K.~E. Strecker, W.~I. McAlexander, G.~B. Partridge,
and R.~G.
  Hulet, Science {\bf 291},  2570  (2001).

\bibitem{Stoof1996a}
H.~T.~C. Stoof, M. Houbiers, C.~A. Sackett, and R.~G. Hulet, Phys.
Rev. Lett.
  {\bf 76},  10  (1996).

\bibitem{Gehm2003a}
M.~E. Gehm, S.~L. Hemmer, S.~R. Granade, K.~M. O'Hara, and J.~E.
Thomas, Phys.
  Rev. A {\bf 68},  011401(R)  (2003).

\bibitem{Diener2004a}
R.~B. Diener and T.-L. Ho, cond-mat/0405174 (2004).

\bibitem{Diener2004b}
R.~B. Diener and T.-L. Ho, cond-mat/0404517 (2004).

\bibitem{Simonucci2005}
S. Simonucci, P. Pieri, and G.~C. Strinati, Europhys. Lett. {\bf
69},  713
  (2005).

\bibitem{Szymanska2005a}
M.~H. Szyma{\'n}ska, K. Goral, T. Kohler, and K. Burnett, Phys. Rev.
A {\bf
  72},  013610  (2005).

\bibitem{Heiselberg2001a}
H. Heiselberg, Phys. Rev. A {\bf 63},  043606  (2001).

\bibitem{Baker1999a}
G.~A. Baker, Phys. Rev. C {\bf 60},  054311  (1999).

\bibitem{Ho2004a}
T.-L. Ho, Phys. Rev. Lett. {\bf 92},  090402  (2004).

\bibitem{O'Hara2002a}
K.~M. O'Hara, S.~L. Hemmer, M.~E. Gehm, S.~R. Granade, and J.~E.
Thomas,
  Science {\bf 298},  2179  (2002).

\bibitem{Menotti2002}
C. Menotti, P. Pedri, and S. Stringari, Phys. Rev. Lett. {\bf 89},
250402
  (2002).

\bibitem{Bourdel2003a}
T. Bourdel {\it et~al.}, Phys. Rev. Lett. {\bf 91},  020402  (2003).

\bibitem{Bartenstein2004a}
M. Bartenstein {\it et~al.}, Phys. Rev. Lett. {\bf 92},  120401
(2004).

\bibitem{Bourdel2004a}
T. Bourdel {\it et~al.}, Phys. Rev. Lett. {\bf 93},  050401  (2004).

\bibitem{Partridge2005a}
G.~B. Partridge, W. Li, R.~I. Kamar, Y. an~Liao, and R.~G. Hulet,
Science {\bf
  311},  503  (2005).

\bibitem{Kinast2005a}
J. Kinast {\it et~al.}, Science {\bf 307},  1296  (2005).

\bibitem{Regal2005c}
C.~A. Regal, M. Greiner, S. Giorgini, M. Holland, and D.~S. Jin,
Phys. Rev.
  Lett. {\bf 95},  250404  (2005).

\bibitem{DeMarco2001c}
B. DeMarco, Ph.D. thesis, University of Colorado, 2001.

\bibitem{Carlson2003a}
J. Carlson, S.~Y. Chang, V.~R. Pandharipande, and K.~E. Schmidt,
Phys. Rev.
  Lett. {\bf 91},  050401  (2003).

\bibitem{Carlson2005a}
J. Carlson and S. Reddy, Phys. Rev. Lett. {\bf 95},  060401  (2005).

\bibitem{Astrakharchik2004a}
G.~E. Astrakharchik, J. Boronat, J. Casulleras, and S. Giorgini,
Phys. Rev.
  Lett. {\bf 93},  200404  (2004).

\bibitem{Astrakharchik2005b}
G.~E. Astrakharchik, J. Boronat, J. Casulleras, and S. Giorgini,
Phys. Rev.
  Lett. {\bf 95},  230405  (2005).

\bibitem{Burovski2006a}
E. Burovski, N. Prokof'ev, B. Svistunov, and M. Troyer, Phys. Rev.
Lett. {\bf
  96},  160402  (2006).

\bibitem{Perali2004b}
A. Perali, P. Pieri, and G.~C. Strinati, Phys. Rev. Lett. {\bf 93},
100404
  (2004).

\bibitem{Hu2006a}
H. Hu, X.-J. Liu, and P.~D. Drummond, Phys. Rev. A {\bf 73},  023617
(2006).

\bibitem{Hu2006b}
H. Hu, X.-J. Liu, and P.~D. Drummond, Europhys. Lett. {\bf 74},  574
(2006).

\bibitem{Liu2005a}
X.-J. Liu and H. Hu, Phys. Rev. A {\bf 72},  063613  (2005).

\bibitem{Bartenstein2004c}
M. Bartenstein {\it et~al.},  in {\em XIX International Conference
on Atomic
  Physics} (American Institute of Physics, Melville, New York, 2004), p.\ 278.

\bibitem{Chen2006a}
Q. Chen, C.~A. Regal, D.~S. Jin, and K. Levin, Phys. Rev. A {\bf
74},  011601(R)
  (2006).

\bibitem{beta_kin}
From the kinetic energy one would expect $E_{kin}/E_{kin}^0
=\frac{3-2\beta}{3\sqrt{1+\beta}}$.

\bibitem{Mackie2005a}
M. Mackie and J. Piilo, Phys. Rev. Lett {\bf 94},  060403  (2005).

\bibitem{Javanainen2005a}
J. Javanainen, M. Kostrun, M. Mackie, and A. Carmichael, Phys. Rev.
Lett {\bf
  95},  110408  (2005).

\bibitem{Bulgac2006a}
A. Bulgac, J.~E. Drut, and P. Magierski, Phys. Rev. Lett. {\bf 96},
090404
  (2006).

\end{thebibliography}


\end{document}